\documentstyle[aps, prl, floats, epsfig, amsmath, amssymb]{revtex}

\title{Many-body position operator in lattice fermionic systems with
  periodic boundary conditions} 
\author{ Bal\'azs Het\'enyi } 
\address{ Institut f\"ur Theoretische Physik, Technische Universit\"at Graz, Petersgasse 16, A-8010 Graz, Austria \\ and \\ Mathematisches
  Institut, Fakult\"at f\"ur Mathematik, Informatik und Statistik, 
  Ludwig Maximilians Universit\"at, Theresienstrasse 39, M\"unchen 80333,
  Germany }

\date{\today}

\begin{document}



\maketitle

\begin{abstract}
  A total position operator $X$ in the position representation is derived for
  lattice fermionic systems with periodic boundary conditions.  The operator
  is shown to be Hermitian, the generator of translations in momentum space,
  and its time derivative is shown to correspond to the total current
  operator in a periodic system.  The operator is such that its moments can
  be calculated up to any order.  To demonstrate its utility finite size
  scaling is applied to the Brinkman-Rice transition as well as metallic and
  insulating Gutzwiller wavefunctions.
\end{abstract}

\begin{center}
{Short title: Periodic lattice position operator}
\end{center}


\newpage

\label{sec:intro}

The position operator and its moments give important information about
localization in quantum systems.  As was shown by Kohn~\cite{Kohn64} metals
and insulators are distinguished by the extent of their localization.  Many
real systems are periodic, and in many model systems periodic boundary
conditions are imposed.  In such cases the Hilbert space which that forms the
domain of operators is restricted hence the position operator is
ill-defined~\cite{Blount62}.  The single particle position operator in the
crystal momentum representation was derived by Blount~\cite{Blount62} and
discussed extensively in the context of band-theory.  In the crystal momentum
representation this operator can be generalized to the many-body
case~\cite{Hine07}.  To calculate the total position in the position
representation Resta~\cite{Resta98,Resta99} suggests to average the quantity
$e^{i\frac{2\pi x}{L}}$.  The expectation value of the total position
operator is then defined as
\begin{equation}
  \langle x \rangle = \frac{L}{2\pi}\mbox{Im}\hspace{.05in}\mbox{ln} \langle \Psi |e^{i\frac{2\pi
      x}{L}}| \Psi \rangle.
\label{eqn:Resta_x}
\end{equation}
Via first order perturbation theory, Resta also shows~\cite{Resta98} that the
time derivative of the polarization operator based on the above definition
gives the total current in the limit $L\rightarrow\infty$.  This idea has
been applied to lattice fermionic systems at half-filling~\cite{Resta99}, and
extended to systems at arbitrary fillings~\cite{Aligia99}.  A related
formalism due to Souza {\it et al.}~\cite{Souza00} based on the cumulant
generating function (of which Eq. (\ref{eqn:Resta_x}) is a special case)
establishes relations between localization and polarization.  

It is important to note that the position operator in this method is
calculated indirectly, by first evaluating the expectation value of
$e^{i\frac{2\pi x}{L}}$.  Eq. (\ref{eqn:Resta_x}) is valid as can be
shown~\cite{Resta99} but the calculation of higher moments is not straight
forward, the spread functional suggested by Resta and Sorella~\cite{Resta99}
(based on eq. (\ref{eqn:Resta_x})) is valid in the thermodynamic limit.

Here it is shown that a total position operator for a lattice fermionic
system with periodic boundary conditions can be defined as the generator of
total momentum shifts.  It is also demonstrated that the time derivative of
the total position operator gives the current for a system with any number of
sites (finite $L$).  The total position operator derived below is such that
expectation values of arbitrary powers are readily evaluated, hence an
accurate assessment and finite size scaling of localization is enabled (up to
any desired order).  The utility of the operator is then demonstrated via
variational calculations on the Hubbard
model~\cite{Hubbard63,Kanamori63,Gutzwiller63} based on the Gutzwiller
wavefunction~\cite{Gutzwiller63,Gutzwiller65}.

\label{sec:method}

The derivation of the total position operator is closely related to that of
the total momentum operator in Ref.  \onlinecite{Essler05}.  The class of
models for which the formalism presented below are those used in strongly
correlated systems consisting of site to site hopping terms and some
interaction terms.  An example of a lattice model is the Hubbard Hamiltonian,
\begin{equation}
H = -t \sum_{\langle i,j \rangle \sigma} (c_{i\sigma}^{\dagger}c_{j\sigma} + H.c.)
    +U \sum_{i \sigma} n_{i \uparrow}n_{i \downarrow},
\end{equation}
consisting of $L$ sites.  In the following, the total position operator will
be derived for the one-dimensional Hubbard model.  Generalizations to
higher dimensions and other lattice models will be discussed below.

The real-space (Wannier state) and reciprocal-space
(Bloch state) creation operators are related in the usual way,
\begin{equation}
\tilde{c}_k = \frac{1}{\sqrt{L}}\sum_{j=1}^L e^{i\frac{2\pi k x_j}{L}} c_j,
\end{equation}
where $x_j$ is the position of site $j$.  In order to define a total position
operator we first define a momentum permutation operator as
\begin{equation}
  P_{kl} = 1 - (\tilde{c}^\dagger_k - \tilde{c}^\dagger_l)(\tilde{c}_k - \tilde{c}_l),
\end{equation}
where $\tilde{c}^\dagger_k$ creates a particle in the Bloch state $k$.
A momentum space shift operator can be defined as
\begin{equation}
  {\mathfrak U}_n = P_{n-1n}....P_{12},
\label{eqn:Un}
\end{equation}
with the property that
\begin{equation}
{\mathfrak U}_L  \tilde{c}_k = \left\{ \begin{array}{rl}
  \tilde{c}_{k-1}{\mathfrak U}_L, & k = 2,...,L \\
  \tilde{c}_{L}{\mathfrak U}_L,   & k = 1.
       \end{array} \right.
\end{equation}
For systems with spin-$\frac{1}{2}$ particles we can define the compound
momentum space shift operator as 
\begin{equation}
  {\mathfrak U} = {\mathfrak U}_{L \uparrow} {\mathfrak U}_{L \downarrow},
\end{equation}
with the property
\begin{equation}
  {\mathfrak U} c_{j,\sigma} = e^{i \frac{2\pi x_j}{L}} c_{j,\sigma} {\mathfrak U},
\label{eqn:Uc}
\end{equation}
where $c_{j,\sigma}$ is an annihilation operator for particles at site $x_j$
with spin $\sigma$.

We define the total position operator $X$ through three conditions.  First we
require it to be the generator of total momentum shifts, i.e. 
\begin{equation}
{\mathfrak U} = e^{i \frac{2\pi X}{L}}.
\label{eqn:x_u}
\end{equation}
We also require $X$ to be Hermitian,
\begin{equation}
X = X^\dagger.
\label{eqn:Herm}
\end{equation}
and that the time derivative of $X$ give the total current,
\begin{equation}
e\dot{X} = ie[H,X] = J,
\label{eqn:current_commutator}
\end{equation}
which for the Hubbard model is defined as
\begin{equation}
J = -i e t \sum_{\langle i,j \rangle \sigma} (c_{i\sigma}^{\dagger}c_{j\sigma} 
    - c_{j\sigma}^{\dagger}c_{i\sigma} ).
\label{eqn:current}
\end{equation}

In order to derive the explicit form of $X$ we first define
\begin{equation}
g(\alpha) = \sum_{x=0}^{L-1} i e^{-i\frac{2\pi x\alpha}{L}},
\end{equation}
which can be evaluated via the geometric sum formula to give
\begin{equation}
g(\alpha) = i\frac{1-e^{-i2\pi\alpha}}{1-e^{-i\frac{2\pi \alpha}{L}}}.
\end{equation}
We can take the derivative of $g(\alpha)$ at some integer value $m$ for
$\alpha$,
\begin{equation}
g'(m) = \frac{2\pi}{L} \sum_{x=0}^{L-1} x e^{-i\frac{2\pi xm}{L}}.
\label{eqn:g'm}
\end{equation}
Inverting the Fourier series, we can obtain an expression for the position
$x$ valid for $x=0,...,L-1$,
\begin{equation}
x = \frac{1}{2 \pi}\sum_{k=1}^{L} g'(m) e^{i\frac{2\pi xm}{L}}.
\end{equation}
For $m\neq L$, 
\begin{equation}
g'(m)=2\pi/(e^{-i \frac{2\pi m}{L}} - 1),
\end{equation}
and $g'(L)$ can be evaluated from Eq. (\ref{eqn:g'm}) using the arithmetic
sum formula giving $g'(L) = \pi(L-1)$.  Thus, an overall expression for $x$
reads as
\begin{equation}
  x = \sum_{m=1}^{L-1} \left( \frac{1}{2} + \frac{e^{-i\frac{2\pi x
          m}{L}}}{e^{-i\frac{2\pi m}{L}}-1}\right).
\label{eqn:x_sawtooth}
\end{equation}
The right hand side of Eq. (\ref{eqn:x_sawtooth}) is the sawtooth function
$f(x) = x \mbox{mod} L$.  We propose to take the sawtooth function as the
definition of our position operator.  Based on Eq. (\ref{eqn:x_u}) we write
the total position operator $X$ for a many-particle system as a power series
in the momentum shift operator as
\begin{equation}
  X = \sum_{m=1}^{L-1} \left( \frac{1}{2} +
    \frac{{\mathfrak U}^m}{e^{-i\frac{2\pi m}{L}}-1}\right).
\label{eqn:X}
\end{equation}
It is to be emphasized that $X$ is a genuine many-body operator (as is that
of Resta~\cite{Resta98}).

Having defined our total position operator, we can now test whether it
satisfies the requirements (Eqs. (\ref{eqn:x_u}), (\ref{eqn:Herm}), and
(\ref{eqn:current_commutator})).  Letting $X$ operate on an arbitrary Wannier
state ($|{\bf x}, {\bf \sigma}\rangle =
c_{1,\sigma_1}^\dagger...c_{N,\sigma_N}^\dagger|0\rangle$) for a system gives
the result
\begin{eqnarray}
  X|{\bf x}, {\bf \sigma}\rangle &= &
\sum_{m=1}^{L-1} \left( \frac{1}{2} +
    \frac{e^{i\frac{2\pi m (x_1 + ... + x_N)}{L}}}{e^{-i\frac{2\pi
          m}{L}}-1}\right)|{\bf x}, {\bf \sigma}\rangle \nonumber\\
& =& ((x_1 + ... + x_L)\mbox{mod}L)|{\bf x}, {\bf \sigma}\rangle
\end{eqnarray}
where we have used Eqs. (\ref{eqn:Uc}) and (\ref{eqn:x_sawtooth}).  Since
\begin{equation}
  {\mathfrak U} |{\bf x}, {\bf \sigma}\rangle = e^{i\frac{2\pi (x_1 + ... + x_N)}{L}} |{\bf x}, {\bf \sigma}\rangle,
\end{equation}
Eq. (\ref{eqn:x_u}) follows.  Hermiticity of $X$ follows from the unitarity
of ${\mathfrak U}$ and from the fact that ${\mathfrak U}^L=1$.

To demonstrate that the operator $X$ satisfies the condition in Eq.
(\ref{eqn:current_commutator}), we first note that ${\mathfrak U}$ commutes with the
interaction part of the Hamiltonian.  This can be shown using Eq.
(\ref{eqn:Uc}).  Thus our task consists of evaluating the commutator $[T,X]$,
$T$ denoting the kinetic part of the Hubbard Hamiltonian.  We first define an
operator
\begin{equation}
  Y = \sum_{m=1}^{L} \frac{{\mathfrak U}^m}{e^{-i\frac{2\pi m}{L}}-1}.
\label{eqn:funny_X}
\end{equation}
The last term in the sum is divergent.  However, below we show that this
divergence disappears for the commutator $[T,Y]$.

We first evaluate the commutator 
\begin{equation} 
[T,Y] = \sum_{m=1}^{L} \frac{[T,{\mathfrak U}^m]}{e^{-i\frac{2\pi
        m}{L}}-1}.
\label{eqn:comtx}
\end{equation}
We split the kinetic energy in two parts as
\begin{eqnarray}
A &=& -t\sum_{\langle i,j \rangle \sigma} c_{i\sigma}^{\dagger}c_{j\sigma}
\nonumber \\
A^\dagger &=& -t\sum_{\langle i,j \rangle \sigma} c_{j\sigma}^{\dagger}c_{i\sigma},
\end{eqnarray}
thus we can rewrite Eq. (\ref{eqn:comtx}) as
\begin{equation}
[T,Y] = \sum_{m=1}^{L}
  \frac{[A,{\mathfrak U}^m]+[A^\dagger,{\mathfrak U}^m]}{e^{-i\frac{2\pi m}{L}}-1}.
\label{eqn:comtx2}
\end{equation}
Each commutator in Eq. (\ref{eqn:comtx2}) can be evaluated using Eq.
(\ref{eqn:Uc}).  We obtain
\begin{eqnarray} 
[A,{\mathfrak U}^m] &=& (e^{-i\frac{2\pi m}{L}}-1){\mathfrak U}^m A
  \nonumber \\
  \mbox{[}A^\dagger,{\mathfrak U}^m\mbox{]} &=& (1-e^{-i\frac{2\pi m}{L}})A^\dagger {\mathfrak U}^m,
\end{eqnarray}
giving a new expression for the commutator
\begin{equation}
[T,Y] = \sum_{m=1}^{L}
  {\mathfrak U}^m A - A^\dagger {\mathfrak U}^m.
\label{eqn:comtx3}
\end{equation}
We now substitute the condition in Eq. (\ref{eqn:x_u}) and we obtain
\begin{equation}
[T,Y] = \sum_{m=1}^{L}
  e^{i\frac{2\pi Xm}{L}} A - A^\dagger e^{i\frac{2\pi Xm}{L}}.
\end{equation}
It is easily seen that this commutator is zero, since $X$ operating on a
Wannier state gives an integer and
\begin{equation}
\sum_{m=1}^{L} e^{i\frac{2\pi Xm}{L}} = 0.
\end{equation}
On the other hand, using the same reasoning we used to arrive at Eq.
(\ref{eqn:comtx3}) it can be shown that
\begin{equation}
[T,X] = \sum_{m=1}^{L-1}
  {\mathfrak U}^m A - A^\dagger {\mathfrak U}^m,
\label{eqn:comtx4}
\end{equation}
hence, from Eq. (\ref{eqn:comtx3}) we see that
\begin{equation}
[T,X] =   A^\dagger - A, 
\label{eqn:comtx5}
\end{equation}
since ${\mathfrak U}^L=1$.  From Eq. (\ref{eqn:comtx5}) the expression for the current
(Eq. (\ref{eqn:current})) follows straightforward.

The total position operator $X$ derived above can be generalized to many
dimensions as follows.  In higher dimensions the operator becomes a vector
operator.  The generalization of the above derivation has to be based on a
generalized total momentum shift operator consisting of the product of all
one-dimensional momentum shift operators in a particular direction.  For
example, for a three dimensional system with dimensions $x,y,z$ a total
momentum shift operator for the $x$ direction (spinless case) would consist
of the product of all one dimensional momentum shift operators
\begin{equation}
 {\mathfrak W}_{L,x} = \prod_{y,z} {\mathfrak U}_{L,x}^{(y,z)},
\end{equation}
where ${\mathfrak U}_{L,x}^{(y,z)}$ denotes the total momentum shift operator
in the $x$-direction for a given set of coordinates $y,z$
(Eq. (\ref{eqn:Un})).  Such an operator satisfies the commutation relation
\begin{equation}
{\mathfrak W}_{L,x}  \tilde{c}_{k_x,k_y,k_z} = \left\{ \begin{array}{rl}
  \tilde{c}_{k_x-1,k_y,k_z}{\mathfrak W}_{L,x}, & k_x = 2,...,L;k_y,k_z=1,...,L \\
  \tilde{c}_{L,k_y,k_z}{\mathfrak W}_{L,x},   & k_x = 1;k_y,k_z=1,...,L.
       \end{array} \right.
\end{equation}
Subsequent construction of a total position operator for a three dimensional
systems follows the same steps as the one-dimensional case.  The total
momentum shift operator for a spin-$\frac{1}{2}$ system can be written as
\begin{equation}
{ \mathfrak  W}_{i} = {\mathfrak W}_{L,i,\uparrow} {\mathfrak W}_{L,i,\downarrow},
\end{equation}
where ${\mathfrak W}_i$ is a vector operator, and $i=x,y,z$.  A particular
component of the total position operator can then be written as 
\begin{equation}
  R_i = \sum_{m=1}^{L-1} \left( \frac{1}{2} +
    \frac{{\mathfrak W}_i^m}{e^{-i\frac{2\pi m}{L}}-1}\right).
\end{equation}
The commutator of operator $R_i$ will give the current in the $i$ direction.
This is a consequence of the fact that the operator ${\mathfrak W}_i$
commutes with the hoppings in directions other than $i$ included in the
Hubbard Hamiltonian.

Extensions of the Hubbard model can also be handled.  More complex
interaction types (nearest neighbor, etc.) follow the same derivation as
above, as the expression for the current does not change in this case.  For
more complex hoppings the expression for the current is modified to include
the new hoppings, but the derivation presented above is still valid.

For impurity models~\cite{Mahan00,Imada98} the strategy of derivation of a
total position operator is modified slightly.  For example, the
one-dimensional periodic Anderson model, in which each site contains a set of
localized $f$-orbitals, can be written as
\begin{equation}
H = -t \sum_{\langle i,j \rangle \sigma} (c_{i\sigma}^{\dagger}c_{j\sigma} +
H.c.) + E_f \sum_{i,l,\sigma} n_f(i,l,\sigma) + \frac{1}{2} \sum_{i}
 \sum_{l,\sigma \neq l',\sigma'} U(l,l') n_f(i,l,\sigma) n_f(i,l',\sigma') +
 H',
\label{eqn:PAM1}
\end{equation}
with
\begin{equation}
H' = \sum_{i,l,\sigma}\{V_l f^\dagger_{i,l,\sigma}c_{i,\sigma} + H.c.\}.
\label{eqn:PAM2}
\end{equation}
In Eqs. (\ref{eqn:PAM1}) and (\ref{eqn:PAM2})
$n_f(i,l,\sigma)$($f^\dagger_{i,l,\sigma}$) denotes the density(creation
operator) of $f$-orbital with label $l$ at site $i$ and with spin $\sigma$.
Each lattice site contains a set of $f$ orbitals, but there are no inter-site
hoppings between the localized $f$-orbitals on different sites.  As a
consequence the current operator is the same as that of the Hubbard model,
inspite of the fact that the charge density includes the $f$-orbital
terms~\cite{Baeriswyl87}.  One could construct a total position operator
which does not include impurity orbitals, and has the same form as $X$
derived above (only electrons in the conduction band enter the definition).
As conduction takes place only on the standard lattice sites, not the ones
associated with the $f$-orbitals, such an approach may in some cases be
sufficient to characterize localization phenomena associated with
metal-insulator transitions.  However it is also possible to construct a
total position operator valid for a system with the periodic Anderson
Hamiltonian.

To do this one has to consider the $f$-orbitals as separate lattices, and
construct a total momentum shift operator for each set of $f$-orbitals
localized on different lattice sites.  One can construct an operator
\begin{equation}
  {\mathfrak V}_L^{(l)} = Q_{L-1L}^{(l)}....Q_{12}^{(l)},
\label{eqn:Vn}
\end{equation}
where
\begin{equation}
  Q_{jk}^{(l)} = 1 - (\tilde{f}^\dagger_{j,l} - \tilde{f}^\dagger_{k,l})(\tilde{f}_{j,l} - \tilde{f}_{k,l}).
\end{equation}
$\tilde{f}_{j,l}$ denotes the Fourier transform of the annihilation operators
of a particular $f$-orbital,
\begin{equation}
  \tilde{f}_{k,l} = \frac{1}{\sqrt{L}}\sum_{j=1}^L e^{i\frac{2\pi k x_j}{L}} f_{j,l}.
\end{equation}
The operator in Eq. (\ref{eqn:Vn}) satisfies the property
\begin{equation}
{\mathfrak V}_L^{(l)}  \tilde{f}_k = \left\{ \begin{array}{rl}
  \tilde{f}_{k-1,l}{\mathfrak V}_L^{(l)}, & k = 2,...,L \\
  \tilde{f}_{L,l}{\mathfrak V}_L^{(l)},   & k = 1.
       \end{array} \right.
\end{equation}
Thus a total momentum shift operator can be constructed as
\begin{equation}
{\mathfrak Z} = {\mathfrak U} \prod_{l}{\mathfrak V}^{(l)},
\end{equation}
where
\begin{equation}
{\mathfrak V}^{(l)} = {\mathfrak V}_{L,\uparrow}^{(l)} {\mathfrak V}_{L,\downarrow}^{(l)}.
\end{equation}
The total momentum shift operator $\mathfrak{Z}$ can be used to construct a
total position operator 
\begin{equation}
  X_{PAM} = \sum_{m=1}^{L-1} \left( \frac{1}{2} +
    \frac{{\mathfrak Z}^m}{e^{-i\frac{2\pi m}{L}}-1}\right).
\end{equation}
The operator $X_{PAM}$ includes the positions of electrons in impurity
orbitals as well as those in the conduction band.  To prove that it satisfies
the three required conditions proceeds as before.  Proving that the
time-derivative of the position operator is equal to the current is
simplified by the fact that the operators $\mathfrak{V}^m$ commute with the
periodic Anderson Hamiltonian.  This is another consequence of the fact that
there are no hoppings between $f$-orbitals positioned on different sites.
Hence all that needs to be proven is that the commutator corresponding to
$\mathfrak{U}$ gives the current operator corresponding to that of the
Hubbard model~\cite{Baeriswyl87}.  This was already shown above.

The operator $X$ is well defined in the occupation number representation and
it and its moments can thus be calculated in practical situations.  Here we
demonstrate the utility of the operator $X$ by calculating the moments and
performing finite size scaling for the Gutzwiller approximate solution of the
Hubbard model at half-filling.  The Gutzwiller wavefunction (GWF) has the
form
\begin{equation}
|\Psi\rangle = \exp\left(-\gamma \sum_i n_{i\uparrow}
n_{i\downarrow}\right)|\Psi_0\rangle.
\label{eqn:GWF}
\end{equation}
where $|\Psi_0\rangle$ is a noninteracting wavefunction, and $\gamma$ is a
variational parameter which projects out double occupations.  Most often
$|\Psi_0\rangle$ is the Fermi sea.  In this case the exact solution in
one~\cite{Metzner87,Metzner88} and infinite
dimensions~\cite{Metzner89,Metzner90} are available.  At half-filling the
former is metallic for finite $U$, in contradiction with the exact
solution~\cite{Lieb68}.  An approximate solution to the GWF due to Gutzwiller
(GA) results in the Brinkman-Rice metal insulator
transition~\cite{Gutzwiller65,Brinkman70,Vollhardt84}.  In finite dimensions
the GA is only approximate, however in infinite dimensions it correponds to
the exact solution~\cite{Metzner89,Metzner90}.  In a one-dimensional system
the Brinkman-Rice transition is known to occur at $U_c\approx10$.  If
$|\Psi_0\rangle$ is a non-interacting antiferromagnetic wavefunction the
Gutzwiller wavefunction can be made insulating~\cite{Metzner89b}.  In the
following, to assess the localization accompanying the metal-insulator
transition we calculate the quantity
\begin{equation}
\chi_4 = \frac{\sqrt{\langle X^4 \rangle - \langle X^2 \rangle
  \langle X^2 \rangle}}{L^2},
\label{eqn:chi}
\end{equation}
via quantum Monte Carlo methods~\cite{Yokoyama86,Hetenyi09}.

\begin{figure}[htp]
\vspace{1cm}
\psfig{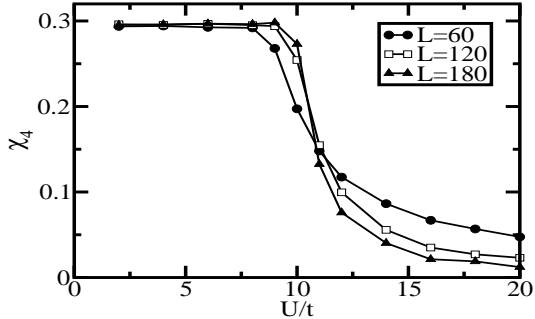}
\vspace{1cm}
\caption{$\chi_4$ (defined in Eq. (\ref{eqn:chi})) for the Hubbard model using
  the Gutzwiller wavefunction evaluated in the Gutzwiller approximation
  scheme.  The Brinkman-Rice transition is known to occur at $U_c\approx10$.}
\label{fig:std}
\end{figure}

In Fig. \ref{fig:std} $\chi_4$ as a function of the Hubbard interaction
strength for three different system sizes is presented.  A transition at
$U_c\approx10$ is clearly visible from the simultaneous drop of all three
curves.  For large $U$ ($U\geq11$) the largest(smallest) system shows the
smallest(largest) value of the fourth moment, which is the tendency one
expects for the insulating state.  (The same behaviour was found for the
square-root of the second order deviation.)  These results coincide with what
is known about the Brinkman-Rice transition being a localization
transition~\cite{Vollhardt84}.
\begin{figure}[htp]
\vspace{1cm}
\psfig{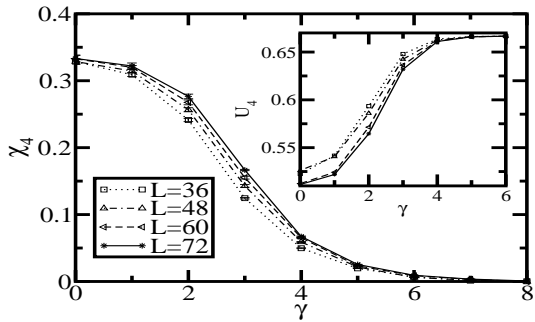}
\vspace{1cm}
\caption{Size dependence of $\chi_4$ for a metallic Gutzwiller wavefunction.
  The inset shows the size dependence of the fourth order Binder cumulant.}
\label{fig:gutzmet}
\end{figure}

\begin{figure}[htp]
\vspace{1cm}
\psfig{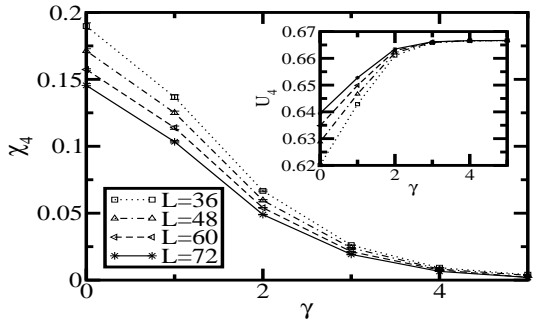}
\vspace{1cm}
\caption{Size dependence of $\chi_4$ for an insulating Gutzwiller
  wavefunction.  The inset shows the size dependence of the fourth order
  Binder cumulant.  }
\label{fig:gutzins}
\end{figure}

In Figs. \ref{fig:gutzmet} and \ref{fig:gutzins} a metallic and an insulating
wavefunction are compared.  For the former the noninteracting wavefunctions
(ground state of the $U=0$ system) was used in place of $|\Psi_0\rangle$ in
Eq. (\ref{eqn:GWF}).  For the insulating wavefunction an antiferromagnetic
solution was used with a magnetization of $m=0.33333$.  The size dependence
of the quantity $\chi_4$ is clearly sensitive to whether the system is
metallic or insulating: as the variational parameter $\gamma$ is increased
$\chi_4$ decreases in both cases, but the size dependence of $\chi_4$ is
opposite between the two cases.  The metallic state (Fig. \ref{fig:gutzmet})
shows an increase in delocalization with system size, whereas in the
insulating state (Fig. \ref{fig:gutzins}) the larger system is more
localized.  The insets in Figs. \ref{fig:gutzmet} and \ref{fig:gutzins} show
the value of the fourth order Binder
cumulant~\cite{Binder81,Binder87,Binder92} defined as
\begin{equation}
  U_4 = 1 - \frac{\langle X^4 \rangle}{3\langle X^2 \rangle \langle X^2 \rangle},
\end{equation}
a quantity used in the finite size scaling~\cite{Fisher72} of phase
transitions.  $U_4$ approaches a value of two-thirds in the case of perfect
localization.  Again, total order (localization) is approached by both the
metallic and insulating wavefunctions, but the size dependence is the
opposite between the two cases, with the larger system closer to the limiting
value of two-thirds for the insulating wavefunction (hence more localized).

In this paper a total position operator was derived for lattice models.  The
operator satisfies three crucial criteria: it is the generator of total
momentum shifts, it is Hermitian, and its time derivative corresponds to the
total current operator.  The form of the operator is such that the average
total position and its moments can be readily calculated.  Hence Binder
cumulants used in finite size scaling can also be evaluated.  The sensitivity
of such moments and cumulants was also demonstrated by investigating their
size dependence in the Brinkman-Rice transition, and metallic and insulating
Gutzwiller wavefunctions.

Part of this work was performed at the Institut f\"ur Theoretische
Physik at TU-Graz under FWF (F\"orderung der wissenschaftlichen
Forschung) grant number P21240-N16.  Part of this work was performed
under the HPC-EUROPA2 project (project number 228398).  Helpful
discussions with H.~G. Evertz are gratefully acknowledged.


\end{document}